\documentclass[aps,prb,twocolumn,superscriptaddress,floatfix]{revtex4}
\usepackage{amsmath,amssymb,graphicx,graphics,color}
\usepackage{dcolumn}   
\usepackage{bm}   
\usepackage{epstopdf}
\usepackage[T1]{fontenc}
\usepackage{calligra}

\begin{document}

\title{Current fidelity susceptibility and conductivity in one-dimensional
  lattice models \\ 
with open and periodic boundary conditions}

\author{S. Greschner}
\affiliation{Institut f\"ur Theoretische Physik, Leibniz Universit\"at Hannover, 
30167~Hannover, Germany}

\author {A. K. Kolezhuk}
\affiliation{Institute of High Technologies, 
Taras Shevchenko National University of Kiev,  03022 Kiev, Ukraine}
\affiliation{Institute of Magnetism, National Academy of Sciences 
and Ministry of Education,  03142 Kiev, Ukraine}
 
\author {T. Vekua}
\affiliation{Institut f\"ur Theoretische Physik, Leibniz Universit\"at Hannover, 
30167~Hannover, Germany}

\begin{abstract}
We study, both numerically and analytically, the finite size scaling of the
fidelity susceptibility $\chi_{J}$ with respect to the charge or spin current in
one-dimensional lattice models, and relate it to the low-frequency behavior of
the corresponding conductivity. It is shown that in gapless systems with open
boundary conditions the leading dependence on the system size $L$ stems from the
singular part of the conductivity and is quadratic,
with a universal form $\chi_{J}= {7KL^2 \zeta(3)}/{2\pi^4}$ where $K$ is the
Luttinger liquid parameter. In contrast to that, for periodic boundary
conditions the leading system size dependence is directly connected with the regular
part of the conductivity (giving alternative possibility to study low frequency behavior of the
regular part of conductivity) and is subquadratic, $\chi_{J}\propto L^{\gamma(K)}$, ($\gamma$ being a K dependent constant)  in most situations linear,
$\gamma=1$. For open boundary conditions, we also study another
current-related quantity, the fidelity susceptibility to the lattice tilt  $\chi_{\mathcal{ P}}$ and
show that it scales as the quartic power of the system size, 
$\chi_{\mathcal{ P}}={31KL^4 \zeta(5)}/{8 u^2 \pi^6}$, where $u$ is the sound velocity. 
We comment on the behavior of the current fidelity susceptibility in gapped phases, particularly in the
topologically ordered Haldane state.
\end{abstract}

\pacs{64.60.-i,75.40.Gb,71.10.Pm,72.15.Nj}
\maketitle

\section{Introduction}
\label{sec:intro}

The ground state fidelity susceptibility 
(FS) has established itself as a useful computational tool 
for locating quantum phase transitions in many-body systems
\cite{You+07,VenutiZanardi07,Schwandt+09,Gu10rev}.
For a general Hamiltonian, 
\begin{equation} 
\label{ham-gen} 
 \mathcal{\widehat{H}}(\lambda)=\mathcal{\widehat{H}}_{0}+\lambda \widehat{W}
\end{equation}
with a phase transition driven by
the coupling to a certain  operator $\widehat{W}$, the fidelity\cite{ZanardiPaukovich}
$F(\lambda,\delta\lambda)=\langle \psi_0(\lambda)|
\psi_0(\lambda+\delta\lambda)\rangle$ measures the change in the ground state
wave function $|\psi_{0}(\lambda)\rangle$ with the infinitesimal change of the
coupling $\lambda$, and the fidelity 
susceptibility $\chi_{W}$ with respect to the ``perturbation''
$W$ is defined as \cite{You+07,VenutiZanardi07}
\begin{eqnarray} 
\label{fs-def}
\chi_{W}(\lambda)&=& \lim_{\delta\lambda\to 0} \frac{1 - |F(\lambda,\delta\lambda)|^2}{\delta\lambda^{2}} \nonumber\\
&=&\sum_{n\not=0} \frac{|\langle \psi_{0}(\lambda)| \widehat{W}|
  \psi_{n}(\lambda)\rangle|^{2}}{\big(E_{n}(\lambda)-E_{0}(\lambda)\big)^{2}} ,
\end{eqnarray}
where second equality is derived, in the second order of perturbation
theory \cite{You+07,VenutiZanardi07} assuming that the ground state is unique. Summation in (\ref{fs-def}) is over all eigenstates $|\psi_{n}(\lambda)\rangle$ of the Hamiltonian
 $\mathcal{H}(\lambda)$ with the eigenvalues $E_{n}(\lambda)$, except the
ground state $|\psi_{0}(\lambda)\rangle$.

Typicaly, in thermodynamic limit the FS would diverge at the point $\lambda_{c}$ corresponding to a quantum phase
transition $\chi_{W}\propto L/(\lambda-\lambda_{c})^{\alpha}$ and for systems of a finite size $L$
the analysis of the scaling behavior of 
\begin{equation} 
\label{fs-peak} 
\chi_{W}(\lambda_{c})\propto
L^{\mu}
\end{equation}
 allows one to extract the critical exponent $\nu$ of the correlation
length, $\nu=(\mu-1)/\alpha$ and thus to determine the universality class of the
transition. 

One of the most advanced unbiased
numerical method for analyzing lattice models in reduced spatial dimensions is
the density matrix renormalization group\cite{White,Uli} (DMRG), which is best suited for systems with open boundary
conditions at least along one of the directions. In one dimension (1d), systems with open boundaries 
consisting of $L \sim
10^2$-$10^3$ sites  can be efficiently analyzed by DMRG. Hence, it
is crucial to understand the dependence of the FS on the boundary
conditions. For many types of the ``perturbation'' $W$, the FS depends only weakly on the boundary
conditions  for large systems. 

In the present paper, we show that if $\widehat{W}$ is charge or spin current operator $\widehat{J}$, or the ``polarization'' operator
$\widehat{\mathcal{P}}$ (which physically corresponds to introducing the
external electric field for charged particles, or to tilting the lattice for
neutral particles, or to a magnetic field gradient for spins), the situation is
very special.  We study the current FS in several model systems, including spin
chains, the Hubbard model for spinful fermions, and the Bose-Hubbard model.  It is shown that in gapless 1d systems with open boundary
conditions (o.b.c.) the leading terms in the $L$ dependence are given by
$\chi_{J}\propto KL^{2}$ and $\chi_{\mathcal{P}}\propto KL^{4}/u^{2}$
respectively, where $K$ is the Luttinger liquid parameter, $u$ is the
characteristic ``sound'' velocity, and the numerical prefactors are
universal. We show, by means of relating $\chi_{J}$ and $\chi_{\mathcal{P}}$ to
the behavior of the positive frequency conductivity $\sigma_1(\omega)$, that
those superextensive terms in the FS originate from the low-frequency behavior
of the singular part of the conductivity.
Since those terms are universal, they can mask the diverging part of the FS at a
phase transition point between two gapless regions.

In contrast to that, for gapless systems with periodic boundary conditions
(p.b.c.) the leading system size
dependence of the current FS is linear, $\chi_{J}\propto L$, in a wide range of
the Luttinger liquid parameter $K$, and may change to a subquadratic one,
$\chi_{J}\propto L^{\gamma}$ with $\gamma$ depending on $K$, $ 1\le \gamma <
2$. At Kosterlitz-Thouless (KT) metal-insulator transition point $\chi_{J}\propto (L/\ln{L})^2$.

As a byproduct of this study, we establish the general properties of
the low-frequency behavior of the conductivity in  systems with open and
periodic boundary conditions. We emphasize a crucial difference in the  behavior
of conductivity in systems with p.b.c. and o.b.c., which is responsible for the
peculiar difference in the current FS properties. 

 In gapped phases the current FS is generically extensive,
independent of boundary conditions, $\chi_{J}\propto L$, but it may again acquire the
quadratic system size dependence for topologically ordered states in systems with open boundary conditions,
for example in the singlet ground state of the open Haldane chain, due to non-locally entangled edge
spins.

The structure of the paper is as follows: in Sec.\ \ref{sec:Spin} we consider
the main properties of the current fidelity susceptibility, its relation to the
conductivity, and the dependence on boundary conditions, on the simplest example
of the spin-$\frac{1}{2}$ XXZ chain in its gapless phase (which is equivalent to
nearest-neighbor interacting spinless fermions). We present two ways of
calculating the current FS for open systems: one is based on the free-fermion
picture and involves bosonization arguments for a generalization to the
interacting case, and the other way is based on applying a unitary twist
transformation and reducing the problem to calculating certain integrals of the
(spin) density correlation function. We also present an example
of how the presence of universal quadratic terms in the current FS can hinder
the detection of a phase transitions between two gapless phases of the
1d Bose-Hubbard model.  
In Sec.\ \ref{sec:FHM} we consider the
properties of the current FS in the fermionic Hubbard
model. Sec.\ \ref{sec:gapped} comments on the behavior of the current FS in
gapped phases, and Sec.\ \ref{sec:summary} contains a brief summary. In appendix we provide details of 
bosonization calculations used throughout paper for open chains.

\section{Spin-current fidelity susceptibility and conductivity of spin-$\frac{1}{2}$ XXZ chain}
\label{sec:Spin}

We start by considering a spin-$\frac{1}{2}$ XXZ chain with the additional
Dzyaloshinskii-Moriya (DM) coupling, described by the Hamiltonian
\begin{equation}
\label{Spinchain}
\widehat{H}=\widehat{H}_{\rm XXZ}+\widehat{H}_{\rm DM},
\end{equation}
 with
\begin{eqnarray} 
\label{XXZChain} 
\widehat{H}_{\rm XXZ} &=&J\sum_{l}(S_l^xS_{l+1}^x+S_l^yS_{l+1}^y+\Delta
S_l^zS_{l+1}^z),\nonumber\\
\widehat{H}_{DM} &=& d\widehat{J}, \quad 
\widehat{J}=\sum_{l} J_{l} =\sum_{l}(\vec{ S}_l\times \vec {S}_{l+1})^z,
\end{eqnarray}
where $S^{\alpha}_{l}$ are spin-$\frac{1}{2}$ operators acting at site $l$ of the
chain. In what follows, we set the Planck constant $\hbar$ and the
lattice spacing $a$ to unity and measure energy in units of $J=1$.

We will study the current fidelity susceptibility (CFS) $\chi_{J}(d)$ that
describes the response of the ground state to an infinitesimal change of the DM
coupling $d$.  In the following, we study separately the cases of open and
periodic boundary conditions.  We will use the upper index $o$ and $p$ to
distinguish the FS for those two cases.

Consider first the CFS at $d=0$ (the alternative derivation, valid for finite
$d$ and for arbitrary half-integer spin $S$,
is presented later in Sec.\ \ref{subsec:alt}). At $d=0$,
the quantity $\widehat{J}$ has the meaning of the total spin current, because
the local currents $J_{l}$ satisfy the continuity equation $\partial_t S_l^z=
J_{l}-J_{l+1}$.

It is worthwhile to note, that the CFS is identical to the so-called stiffness FS
\cite{ThesbergSorensen11} $\chi_{\rho}$, which describes the response of the
ground state to a uniform infinitesimal twist $\varphi $ on every link: 
\begin{equation}
\label{unitwist}
\widehat{H} \mapsto \widehat{H}(\varphi)=\!\!\!\sum_{j}\Big[\! (\frac{1}{2}S_j^+S^-_{j+1}e^{i\varphi}+ \text{h.c.})+\Delta
S_j^zS_{j+1}^z \Big].
\end{equation}
The second derivative of the ground state energy
with respect to $\varphi$ defines the spin stiffness $\rho$; for spin-$\frac{1}{2}$ XXZ
chain and the Hubbard model, exact results for $\rho$ are available \cite{ShastrySutherland90}. Note, even though $\chi_J=\chi_{\rho}$ the
second derivative of the ground state energy with respect to $d$ and $\varphi$ differ. Following the results of the second order perturbation theory in Ref.~\onlinecite{ShastrySutherland90},
\begin{eqnarray}
\label{twistenergy}
\frac{1}{L}\frac{\partial^2 E_0 }{ \partial \varphi^2}\Big|_{\varphi=0}& =&\frac{1}{L}\left( <-T_k>- 2 \sum_{n\neq 0} \frac {  | < \psi_{0}| \widehat{J}|
 \psi_{n}> |^2 } {E_{n}-E_{0}} \right)\nonumber\\
&=&  2\rho
\end{eqnarray}
gives twice the spin stifness $\rho$ as already mentioned and $T_k=\sum_{j}[\frac{1}{2}S_j^+S^-_{j+1}+ \text{h.c.}] $ is kinetic energy.
Similarly the second order perturbation theory gives~\cite{Schwandt+09}, 
\begin{equation}
\label{currentenergy}
\frac{\partial^2 E_0 }{ \partial d ^2}\Big|_{d=0}= -2 \sum_{n\not=0} \frac{|< \psi_{0}| \widehat{J}|
 \psi_{n} >|^{2}}{E_{n}-E_{0}} .
\end{equation}
 In particular, for p.b.c. $\partial^2 E_0/ \partial d ^2=0$ for $\Delta=0$ as current commutes with kinetic energy in periodic chains and $\rho=1/\pi$.

For the systems with o.b.c. the uniform twist can be completely absorbed by unitary transformation for any $\Delta$ (equivalently due to the $f-$sum rule for any $\Delta$), 
\begin{equation}
\label{twistopen}
\frac{\partial^2 E_0 }{ \partial \varphi^2}\Big|_{\varphi=0} =0
\end{equation}
and 
\begin{equation}
\frac{\partial^2 E_0 }{ \partial d^2}\Big|_{d=0}=    <T_k>,
\end{equation}
hence ${\partial^2 E_0 }/{ \partial d^2}\neq 0$  even for $\Delta=0$ ( $<T_k>=-1/\pi$ for free case ) as current does not commute with kinetic energy for o.b.c..

Note also that if one performs a twist by $L\varphi\ll 1$ only on one link of a \emph{periodic} chain 
(twisting the boundary conditions) the energies will not change as compared to uniform twist of every link with $\varphi$ angle, but the FS with respect to the twist in one link will be different~\cite{ThesbergSorensen11} from $\chi_{\rho}$ of Eq.~(\ref{unitwist}). The reason is that twisting the
single link (twisting the boundary condition) breaks translational symmetry and thus makes the situation similar to that in the o.b.c. case. As
a result, the response of the ground state wavefunction to the infinitesimal twist on a single link is non-zero
independent of boundary conditions, even for the non-interacting case ($\Delta=0$).

Instead of the spin-$\frac{1}{2}$ XXZ chain with the DM coupling,
described by the Hamiltonian (\ref{Spinchain}), we may have in mind
interacting lattice fermions or hard-core bosons, under the
action of some ``field'' $d$ that couples to the total particle current,
\begin{eqnarray}
\label{eq:FermiorBose}
\widehat{H}&=&  -\frac{1}{2} \sum_j \left [ c_j^\dag c^{\vphantom{\dag}}_{j+1} +
  c_{j+1}^\dag c^{\vphantom{\dag}}_{j} \right ]  
+\Delta \sum_j n_j n_{j+1} \nonumber\\
&-&i\frac{d}{2} \sum_j \left [ c_j^\dag c^{\vphantom{\dag}}_{j+1} 
-  c_{j+1}^\dag c^{\vphantom{\dag}}_{j} \right ],
\end{eqnarray}
which is equivalent to the spin-$\frac{1}{2}$ XXZ chain by the well-known
Jordan-Wigner transformation. At $d=0$ CFS $\chi_J$ defines the response of the ground
state of such a system to the infinitesimal uniform change of current through nearest-neighbor links.

\subsection{Relation between the CFS and the conductivity}
\label{subsec:spin-general}

According to (\ref{fs-def}), the CFS can be written as
\begin{equation}
\label{fs-d0-pert}
\chi_{J}(d=0)= \sum_{n\not=0} \frac{ |\langle \psi_0| \widehat{J}  |\psi_n\rangle|^2} {(E_n-E_0)^2}
\end{equation}
where $E_n$ are the eigenvalues of $\widehat{H}_{\rm XXZ}$ and summation is over all excited states.
Comparing the above expression to the definition 
of the positive frequency real part of the spin current conductivity \cite{Giamarchi},
\begin{eqnarray}
\label{conductivity}
\sigma_{1}(\omega)&\equiv&\left.\text{Re}\,\sigma(\omega)\right|_{\omega>0}
\nonumber\\
&=&\frac{\pi}{L\omega} \sum_{n\not=0}   
|\langle \psi_0| \widehat{J} |\psi_n\rangle|^2 \delta(\omega-(E_n-E_0)),
\end{eqnarray}
one obtains the following relation between the CFS and the integrated conductivity:
\begin{equation}
\label{cfs-cond}
\chi_{J}= \frac{L}{\pi}\int_{0}^{\infty} d \omega   \frac{ \sigma_1(\omega)}{\omega } .
\end{equation}
It is important that for p.b.c. systems the definition (\ref{conductivity}) does
not include the Drude weight term $Ku\delta(\omega)$.  The Drude weight is
concentrated at $\omega=0$, while the sum in Eq.\ (\ref{conductivity}) is over
the energy eigenstates with the lower bound $E_n-E_0\sim 1/L>0$, so it does not
account for the zero mode \cite{ShastrySutherland90} .

In contrast to that, in systems with o.b.c. the total current does not commute with
the Hamiltonian even in the noninteracting case ($\Delta=0$). Its zero mode
vanishes identically (see  Eq.~(\ref{nozeromodes}) in the Appendix), hence the singular part of the conductivity (the
Drude weight term) is included in
$\sigma_1(\omega)$. As we will see below, it is due to this reason that the
finite-size scaling of $\chi_{J}$ is quite different for systems with periodic
and open boundary conditions.

\subsection{Periodic boundary conditions}
\label{subsec:spinper}

For a periodic chain 
\begin{equation}
\label{reg-per}
\sigma^p_1(\omega)= \sigma_{\rm reg}(\omega),
\end{equation}
where $\sigma_{reg}(\omega)$ is a regular part of cunductivity; 
as mentioned above, the total real part of the conductivity (including zero mode) is
\begin{equation}
\label{cond-total-per}
\text{Re}\, \sigma^p(\omega)= {Ku}\delta(\omega)+ \sigma_{\rm reg}(\omega) ,
\end{equation}
where $u=\frac{K}{2K-1}\sin{\frac{\pi}{2K}}$ has the meaning of
the spin-wave velocity of the
XXZ chain. The total conductivity satisfies the $f-$sum rule \cite{Pines,ShastrySutherland90,Essler01,Eric02}
\begin{eqnarray} 
\label{1st-rule} 
 \frac{1}{\pi}\int_0^{\infty} \sigma(\omega) \, d \omega     =
 -\frac{1}{2L}\langle T_k \rangle,
\end{eqnarray}
where $\langle T_{k}\rangle$ is the average kinetic energy  which in
the case of the XXZ chain can be evaluated exactly from the dependence of the ground state energy on anisotropy parameter $\Delta$, 
\begin{equation} 
\label{Tk} 
\langle T_{k}\rangle=  E_{0}(\Delta)-\Delta\partial_{\Delta}E_0(\Delta) .
\end{equation}

\begin{figure}[tb]
\includegraphics[width=5.5cm]{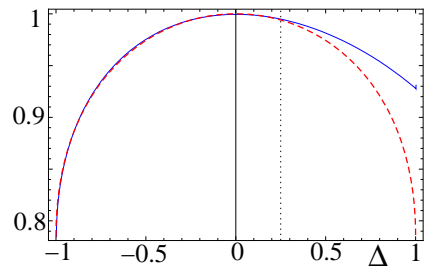}
\caption{Exact results for $-\frac{\pi}{L} \langle T_{k}\rangle $ (continuous line) vs
  $Ku$ (dashed line) in the spin-$\frac{1}{2}$ XXZ chain, as functions of the
  anisotropy $\Delta$ obtained from Bethe Ansatz solution in thermodynamic limit.  The deviation at
  $1/4 \lesssim \Delta$ shows that in this region the regular part of the
  conductivity has to contribute noticeably into the sum rule
  (\ref{1st-rule}). The coefficient in front of the Drude weight of balistic transport is related to stifness as $Ku=\pi \rho$ \cite{ShastrySutherland90}.   }
\label{fig:sumrule}
\end{figure}

One can observe that for the XXZ chain with  $-1<\Delta\lesssim 1/4$ the
product $Ku$ is well approximated by
\begin{equation}
Ku\simeq  \pi \langle -T_k \rangle  /L,
\end{equation}
so the r.h.s. of (\ref{1st-rule}) is, in a rather wide region $-1<\Delta\lesssim
1/4$, well approximated by $Ku/(2\pi)$ and thus in this region the sum rule
(\ref{1st-rule}) is exhausted to a high accuracy by the Drude term (using $\int_0^{\infty} \delta(x)\mathrm{d} x =1/2$).

For weak interaction $\Delta$, a perturbative calculation \cite{Giamarchi91}
yields $ \sigma_{\rm reg}(\omega) \sim \Delta^2 \omega^{8K-5}$. Then for $K>5/8$
(which corresponds to $\Delta<\frac{1+\sqrt{5}}{4}\simeq 0.8$ and strictly speaking is outside the perturbative in $\Delta$ regime) the integral in
(\ref{cfs-cond}) is $O(1)$, so the CFS has a usual extensive dependence on the
system size,
\begin{equation} 
\label{chiJ-largeK} 
\chi_{J}\propto L+ \cdots,\quad K>\frac{5}{8},
\end{equation}
where dots stand for subleading contributions in the system size.
The situation is different for periodic chains with $1/2<K<5/8$, where  
the relation (\ref{cfs-cond}) suggests
the following  non-trivial dependence on the system size: 
\begin{equation}
\label{chiJ-smallK}
\chi^{p}_{J}  \propto  L^{6-8K}  +\cdots, \quad \frac{1}{2} <K <\frac{5}{8}.
\end{equation}
It can be obtained by replacing the lower integration limit in
Eq.\ (\ref{cfs-cond}) by a quantity of the order of $u/L$.

The KT phase transition point $\Delta=1$, where $K=1/2$, must be treated separately, since at
this point the conductivity gets logarithmic corrections \cite{Giamarchi}, $\sigma_{reg}
(\omega)\sim 1/\omega\log^2(\omega)$,
 and hence 
\begin{equation}
\label{chiJ-KT}
\chi^{p}_{J}  \sim  (L/ \ln L )^2 +\cdots,\quad \Delta=1.
\end{equation}

It should be remarked that our results Eqs.\ (\ref{chiJ-largeK})-(\ref{chiJ-KT}) disagree with the conclusions of
Ref.\ \onlinecite{ThesbergSorensen11} who have studied the stiffness FS
$\chi_{\rho}$ that is equal to our current FS $\chi_{J}$ as already mentioned.  First,
Eq.\ (20) and Figs.\ 1 and 2 of Ref.\ \onlinecite{ThesbergSorensen11} create the
impression that the leading contribution to the $\chi_{\rho}$ scales generically
as $L^2$ in \emph{periodic} systems. We believe this is a mistake in the
presentation, since we have perfectly reproduced Figs.\ 1 and 2 of
Ref.\ \onlinecite{ThesbergSorensen11} for the quantity $\chi_{\rho}$ (and not
for $\chi_{\rho}/L$ as it stands in the original paper). Second, in view of the
intimate connection between the CFS and the regular part of conductivity
established by us above, the subleading corrections to the finite-size scaling
of $\chi_{\rho}$, as proposed in Ref.\ \onlinecite{ThesbergSorensen11} and
derived on the basis of analyzing the scaling dimensions of
operators~\cite{VenutiZanardi07} perturbing the Luttinger liquid in the bosonization framework, contradict Giamarchi's result
\cite{Giamarchi91} for the conductivity.

\begin{figure}[tb]
\includegraphics[width=6.8cm]{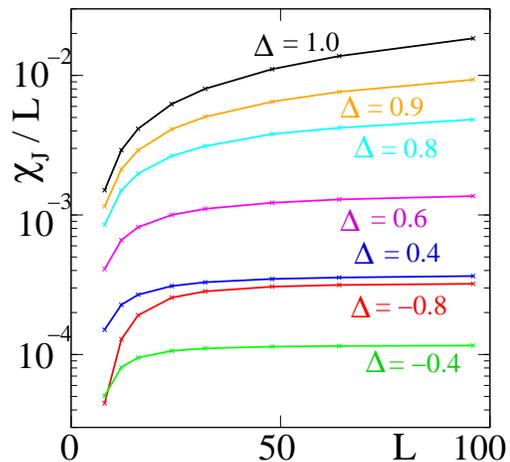}
\caption{The finite-size scaling of the current FS $\chi_J$ in
  spin-$\frac{1}{2}$ XXZ chain with  p.b.c., for different values of the
  anisotropy $\Delta$. In DMRG simulations we take $\delta d=10^{-3}$ and to achieve good accuracy we keep about $m\sim 1000$ states.
Small numerical prefactors of about $\sim 10^{-4}$ to
  $10^{-3}$, observed for $\Delta<0.5$, are related to the fact that the
  contribution from the regular part of the conductivity into the sum rule
  (\ref{1st-rule}) is small.}
\label{fig:Giamarchi}
\end{figure}

Fig.~\ref{fig:Giamarchi} shows the DMRG results for $\chi^{p}_J$ in periodic XXZ
chains with different values of the interaction $\Delta$. For $\Delta \lesssim 0.5$, a
good convergence to the linear scaling $\chi^{p}_J\propto L$ is achieved already
at $L=100$.  However, for larger values of $\Delta$ (especially for $\Delta \to
1$) the value $\chi^{p}_J/L$ does not seem to converge to a constant at
$L\gg 1$, though to make a definitive claim one has to study much larger
system sizes. For the range of $L$ studied here, $L\le 100$, fitting with $\chi^{p}_{J} \sim
(L/\ln L)^2$ as well as with $\chi^{p}_{J} \sim L^{6-8K}$ seems equally
reasonable.

At $SU(2)$ symmetric point $\Delta=1$ it is worthwhile to mention effect of the next-nearest-neighbor antiferromagnetic (as well $SU(2)$ symmetric) interaction $J_2$,
\begin{equation}
\widehat H=\sum_{j}(\vec{ S}_j \vec {S}_{j+1}+J_2\vec{ S}_j \vec {S}_{j+2} ).
\end{equation}

 Observe that due to $J_2$ coupling expression of current operator changes as follows,
\begin{equation}
\label{j1j2current}
 \widehat{J}\to    \frac{i}{2}\sum_j [ S^{+}_jS_{j+1}^- +   J_2 S^{+}_jS_{j+2}^-  ]+ (\text{h.c.})
\end{equation}

At a special point, $J_2=J_{2}^c\simeq 0.241$ (that in thermodynamic limit corresponds to a phase transition point between Luttinger liquid and dimerized phases) the amplitude of the basic (marginal) Umklapp term vanishes
in effective bosonization formulation and hence low frequency behavior of the regular part of conductivity changes to $\sigma_1(\omega) \sim \omega^{8nK-5}$, where  $K=1/2$ due to $SU(2)$ symmetry and $n$ is some integer $n>1$ so that in any case 
$\int \sigma_1(\omega)/\omega \mathrm{d} \omega$ converges at $\omega=0$.
Hence at $J_2=J_2^c$  the CFS $\chi^p_J\sim L$, even though this is a phase transition point between gapless and gapped regions.
This agrees well with the data on Fig. 4 of Ref.\ \onlinecite{ThesbergSorensen11}; namely ratio $\chi^p_{J}/L$ becomes nearly system size independent at $J_2^c$. 
The form of the current operator (\ref{j1j2current}) explains why infinitesimal twist must be the same (and not factor of 2 different) in both nearest-neighbor and next-nearest-neighbor links to observe the flat curve $\chi^p_{J}/L$ vs $L$ at $J_2=J_2^c$.

\subsection{Open boundary conditions}
\label{subsec:spinopen}

\begin{figure}[tb]
\includegraphics[width=4.5cm]{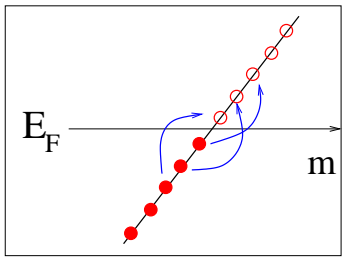}
\caption{An example of the particle-hole excitation with the excitation energy $3\pi u/L$. As
  one can see, there are exactly $\rho(3)=3$ different particle-hole excitations
  (involving the creation of a single particle-hole pair) with that energy. The
  same picture holds for any energy $E_m-E_0=\pi um/L$, hence $\rho(m)=m$.}
\label{fig:Fermisea}
\end{figure}

Let us start our discussion of the CFS for open chains from the non-interacting
case $\Delta=0$ (free spinless fermions or hardcore bosons). At low excitation energies, the
spectrum is approximately linear, $E_{m}-E_{0} \simeq {u \pi m }/{L}$. 
The expression~(\ref{conductivity}), which for o.b.c. represents the
\emph{entire} conductivity, can be rewritten as
 \begin{equation}
\label{cond2}
\sigma^{o}_{1}= \frac {\pi}{\omega L}\sum_{m>0} \rho(m) |\langle \psi_{0}|\widehat{J}|
\psi_m\rangle|^{2}
\delta(\omega-(E_m-E_0))
\end{equation}
where the matrix element of the current (see the Appendix) is
\begin{equation} 
\label{matel-current} 
|\langle \psi_{0}|\widehat{J}|
\psi_m\rangle|=\frac{\big[ 1-(-1)^m\big] u}{m \pi}
\end{equation}
where according to our conventions for free fermions $u(\Delta=0)=1$ and the degeneracy
\begin{equation}
\label{dos}
\rho(m)=m
\end{equation}
is the number of different particle-hole excitations with the same energy
$E_{m}$ as is illustarted in Fig.~\ref{fig:Fermisea} (excited states with more
than one particle-hole pair do not contribute, since they cannot be created by the
current operator from the ground state). Note that the matrix element satisfies
the parity selection rule and is nonzero only for odd $m=2k+1$.

Putting everything together, we obtain the following low-frequency behavior for
the conductivity of free fermions (the XY model, $\Delta=0$) in an open chain:
\begin{eqnarray}
\label{sigma-free}
\sigma_{1}^{o}&=& \frac{\pi u^2}{L \omega} 
\sum^{\infty}_{k=0} \frac{4(2k+1)}{(2k+1)^2 \pi^2}\delta(\omega-(E_{2k+1}-E_0))\nonumber\\
&=&  \sum^{\infty}_{k=0} \frac{4u}{(2k+1)^2
  \pi^2}\delta(\omega-\frac{(2k+1)u\pi}{L}).
\end{eqnarray}
Note that this low-frequency behavior is singular,
$L\sigma_{1}^{o}(\omega)\propto 1/\omega^{2}$ at $\omega\to0$. This is the form, into which the
Drude peak transforms in the open chain.

For the interacting case, we divide the conductivity of the open chain into the
Drude part $D(\omega)$, and the  ``regular'' contribution,
\begin{equation}
\label{drude+regular}
\sigma_1^o(\Delta)=D(\omega)+\sigma^o_{reg}.
\end{equation}
We calculate the Drude part within the Luttinger liquid (LL) approximation
(i.e., we neglect umklapp processes). For the interacting case, the expression
for the current operator does not change, since the interaction commutes with
the local density operator, but the matrix elements of the current
Eq.~(\ref{matel-current}) do change. To separate the Drude contribution, we
estimate the matrix element of the current for the interacting case. Introducing
a bosonic field $\phi$ and its conjugate momentum $\Pi$, which satisfy the
commutation relations $[\phi(x),\Pi(y)]=i\delta(x-y)$, we get the following
Gaussian model as the effective bosonic Hamiltonian of free fermions:
\begin{equation}
\widehat{H}(\Delta=0)= \frac{u(\Delta=0)}{2}\int_0^L dx \big[ (\partial_x \phi)^2+\Pi^2 \big].
\end{equation}
In the LL approach, the presence of the interaction $\Delta$ leads simply to the
rescaling of the bosonic field $\phi \to \tilde \phi= \sqrt{K} \phi$, having the
following rescaling effect on the current:
\begin{equation*}
\widehat{J} \to \int_0^L \frac{\partial_t \phi}{\sqrt{\pi}} \, d x = \sqrt{K}
\int_0^L   \frac{\partial_t \tilde \phi}{\sqrt{\pi}} \, d x = \sqrt{K}u \int_0^L  \frac{ \tilde \Pi}{\sqrt{\pi}}\, d x 
\end{equation*}
and the effective LL Hamiltonian of interacting fermions is 
\begin{eqnarray*}
\widehat{H}_{XXZ}&=&\widehat{H}(\Delta=0)+\Delta\sum_i S^z_lS^z_{l+1}\nonumber\\
& \mapsto&
\widehat{H}_{LL}=\frac{u}{2}\int_0^L \big[ (\partial_x \tilde \phi)^2+\tilde \Pi^2 \big] \, dx .
\end{eqnarray*}
The matrix element of $\int_0^L \tilde \Pi\mathrm{d} x$ between the ground state
and excited states is calculated in the Appendix.

Thus, the effect of interactions on the Drude weight (the singular contribution)
boils down to  rescaling of the matrix elements of the total current for the
free case Eq.~(\ref{matel-current}) 
by the  factor $\sqrt{K}$, and of course the sound velocity $u$ is also renormalized
by the interaction:
\begin{equation}
\label{sigma-int2}
D(\omega)=  \frac{\pi}{L \omega} \sum^{\infty}_{k=0} \frac{4Ku^2}{(2k+1) \pi^2} \delta(\omega-\frac{(2k+1)u\pi}{L}).
\end{equation}
The integral of the Drude part $D(\omega)$ is exactly equal to that of the Drude weight in a periodic
chain:
\begin{equation}
 \int_{0}^{\infty} D(\omega) \, d \omega =\sum_K\frac{4Ku}{(2k+1)^2\pi^2}=\frac{Ku}{2}. 
\end{equation}
Needless to say, as in periodic chains, the singular part $D(\omega)$ almost exhausts the sum rule (\ref{1st-rule}) for $\Delta\lesssim 1/4$.
The regular part of the conductivity in open and periodic chains can, generally
speaking, be different, but the sum rule  requires that 
\[
\lim_{L\to\infty }\frac{1}{L}\int_{0}^{\infty} [\sigma^o_{\rm
    reg}-\sigma_{\rm reg}]\, d\omega \to 0.
\]

Importantly, the leading size dependence of the CFS comes from the singular part
$D(\omega)$, and thus in an open chain the CFS
scales quadratically with the system size:
\begin{equation}
\label{chiJ-L2}
\chi^{o}_{J} =\frac{L}{\pi}\!\int_{0}^{\infty} \!\!\!\!\! d \omega 
\frac{D(\omega)}{\omega}  
\!= \!\!\sum_{k=0} \frac{4K L^2}{ \pi^4 (2k+1)^3}
\!=\!\frac{7\zeta(3)}{2\pi^4}KL^2,
\end{equation}
where $\zeta(x)$ is the Riemann zeta-function. 
This result prominently illustrates the difference between the periodic and open
chains, and shows that one has to be careful when applying the CFS for detecting
phase transitions: unless the transition involves some divergences in the
current correlators, the leading contribution (\ref{chiJ-L2}) will be ``blind''
to it, so the divergence of $\chi_{J}$ at the phase transition will be hidden in
the subleading terms (usually, the exponent $\mu$ that determines the finite-size scaling of
the divergent part of the FS at the phase transition (see Eq.~(\ref{fs-peak}) is
some number between $1$ and $2$). 

\begin{figure}[tb]
\includegraphics[width=6.5cm]{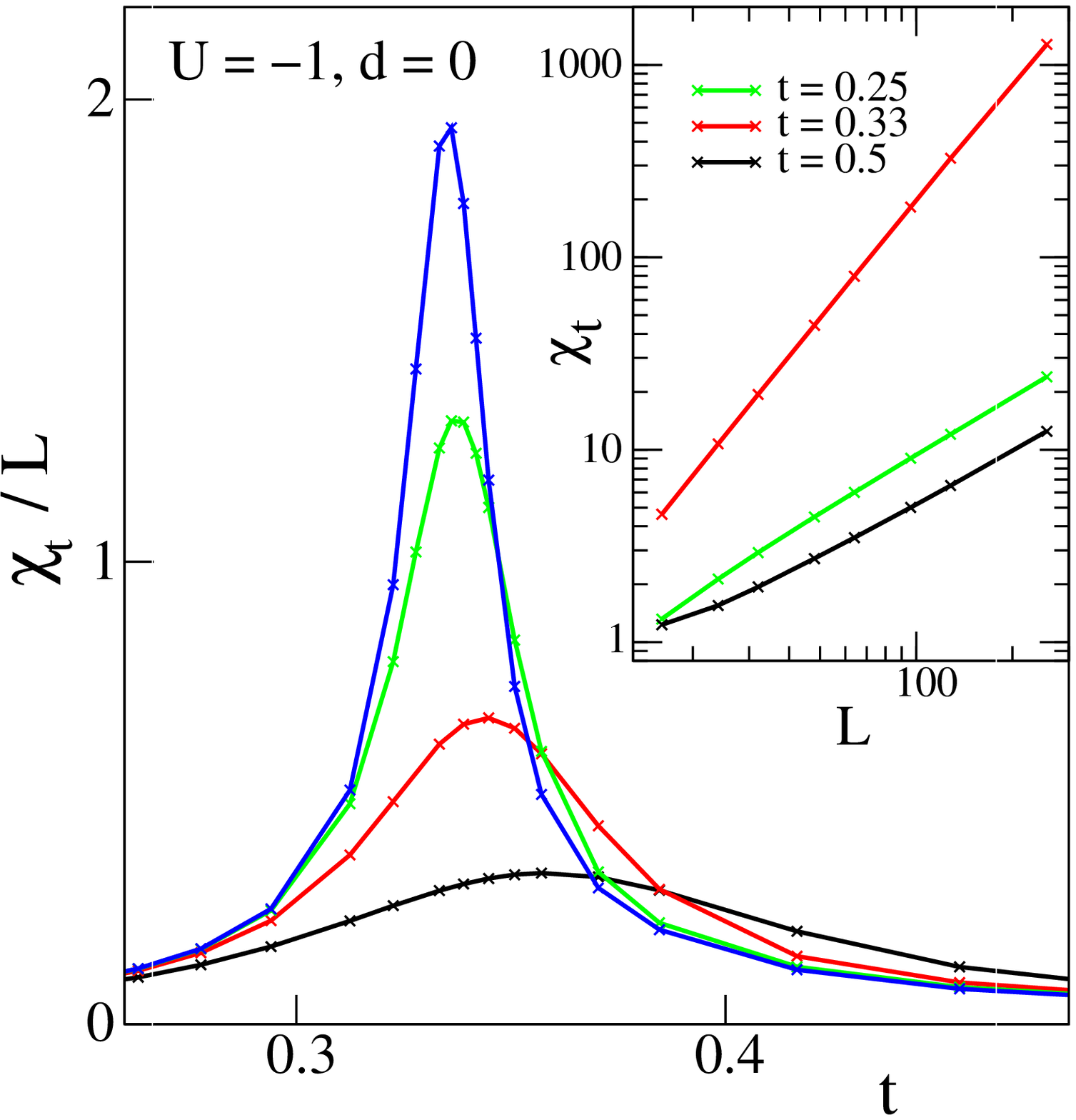} 
\includegraphics[width=6.5cm]{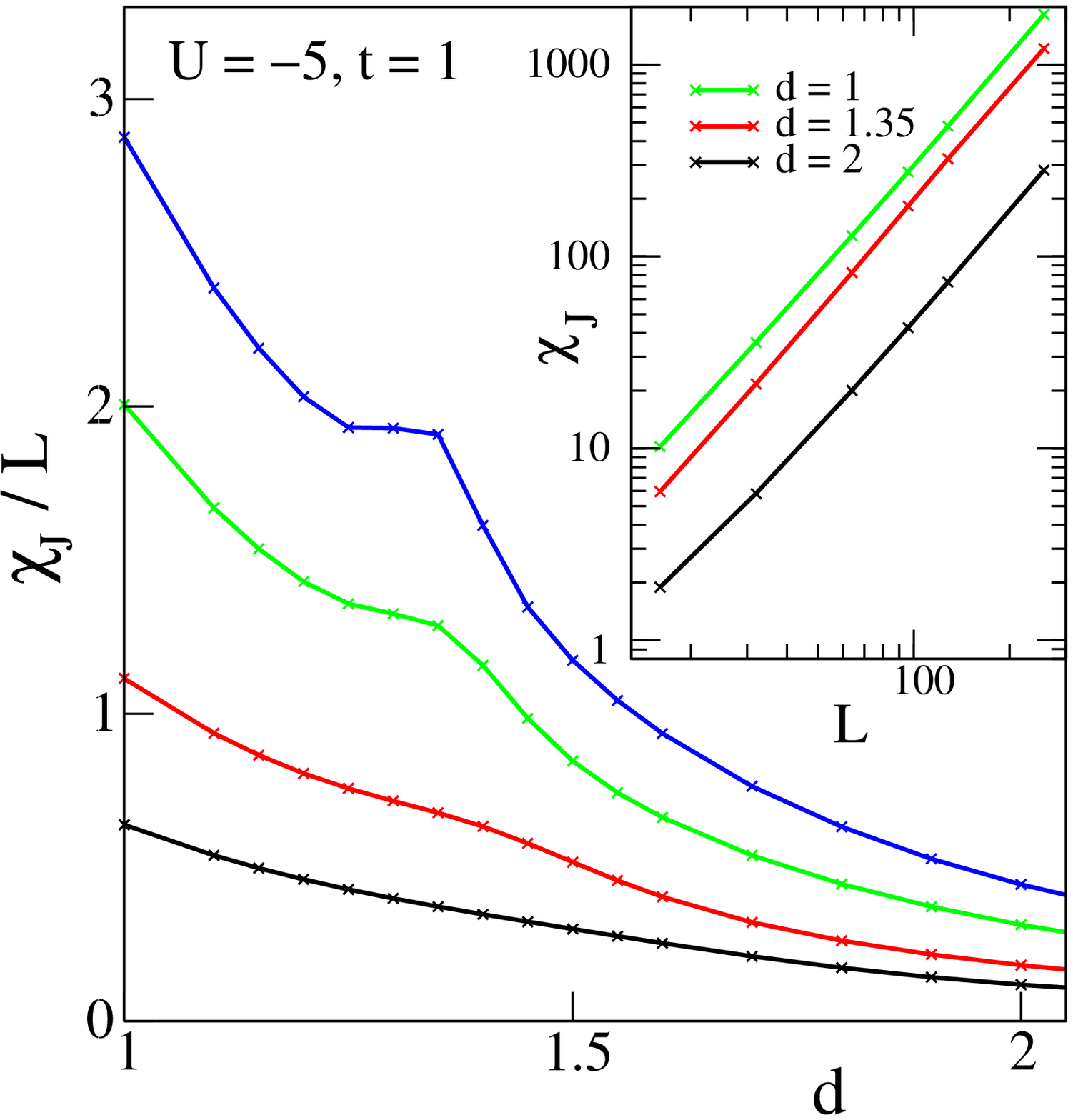}
\caption{ (Color online). Two different fidelity susceptibilities in the
  Bose-Hubbard model with o.b.c. (\ref{eq:H-BH}): the FS $\chi_t$ with respect to the
  hopping $t$ (upper panel) and the current FS $\chi_J$ (lower panel), for open
  chains with $L=16, 32,64$ and $96$ sites (refers to the curves from bottom to
  top). If  $\chi_t$ reveals nicely underlying Ising phase transition between pair and single particle condensates, looking at $\chi_J$, the finite-size scaling of the peak is masked by its wings: both of them scale as $\sim L^2$. }
\label{fig:BoseHubbard}
\end{figure}

We illustrate such ``masking''   on the example of the attractive
single-component Bose-Hubbard model with the additional
 3-body occupation constraint \cite{Greschner}: 
\begin{eqnarray}
\label{eq:H-BH}
H&=&  -\frac{t}{2} \sum_j \left [ b_j^\dag b^{\vphantom{\dag}}_{j+1} +
  b_{j+1}^\dag b^{\vphantom{\dag}}_{j} \right ]
-\frac{id}{2} \sum_j \left [b_j^\dag b^{\vphantom{\dag}}_{j+1} 
-  b_{j+1}^\dag b^{\vphantom{\dag}}_{j} \right ]  \nonumber\\
&+&\frac{U}{2}\sum_j n_j(n_j-1)+U_3\sum_j n_j(n_j-1)(n_j-2),
\end{eqnarray}
where $b_j^\dag$, $b_j$ are the bosonic creation/annihilation operators of
particles at site $j$, $n_j=b_j^\dag b_j$, and the three-body coupling constant
$U_3\to \infty$ forbids sites with more than double occupancy.
Fig.\ \ref{fig:BoseHubbard} presents the DMRG results for the FS study of the
Ising phase transition between the single-particle superfluid and pair
superfluid states (see the phase diagram in Ref.\ \onlinecite{Greschner}).  
The transition is easily detected by looking at the FS with respect to the hopping part (changing $t$), but when
it is studied by looking at the current FS (i.e., the parameter $d$
is changed), it is masked for
chains with o.b.c. as is seen in the lower panel of Fig.\ \ref{fig:BoseHubbard}.

\subsection{Alternative derivation of the CFS scaling  for open boundary conditions}
\label{subsec:alt}

The CFS behavior in spin-$S$ XXZ chain with arbitrary half-integer $S$
can be analyzed with the help of a
different approach, valid at any $d$ as well as at finite
magnetization $M$ (i.e., in presence of some external magnetic
field). Consider a unitary transformation defined by the twist
operator
\begin{equation} 
\label{twist} 
\widehat{U}[\phi(d)] =e^{i\phi(d) \widehat{\mathcal{P}}}, \quad \widehat{\mathcal{P}} = \sum_j j S_j^z, 
\end{equation}
where $\phi(d)=\arctan({d})$ and $\widehat{\mathcal{P}}$ is the
``polarization'' operator, or the ``spin center of mass''. Applied to the Hamiltonian Eq.~(\ref{Spinchain}), 
it removes the DM interaction, for the price of changing the
anisotropy. Performing two such transformations, $U(d)$ and $U(d+\delta d)$,  
one can transform the fidelity $F(d,\delta d)$ into the matrix element of the
form
\begin{equation} 
\label{fidelity-twisted} 
F(d,\delta d)=\big\langle \psi_{0}(M,\widetilde{\Delta}_{d}) \big| 
e^{i\big[\phi(d+\delta d)-\phi(d)\big]\widehat{\mathcal{P}}} 
\big| \psi_{0}(M,\widetilde{\Delta}_{d+\delta d})\big\rangle,
\end{equation}
where $\widetilde{\Delta}_{d}=\Delta/\sqrt{1+d^{2}}$. Expanding the
fidelity up to quadratic terms in $\delta d$, one obtains
\begin{eqnarray} 
\label{fidelity-twisted-expand} 
F(d,\delta d)&\simeq& 1-\frac{1}{2}\delta\Delta^{2} \chi_{\Delta}\nonumber\\
&+&i\delta\phi \langle \psi_{0}(M,\widetilde{\Delta}_{d})|
\widehat{\mathcal{P}} | \psi_{0}(M,\widetilde{\Delta}_{d+\delta
  d})\rangle \nonumber\\
&-&\frac{1}{2}\delta\phi^{2}\langle \psi_{0}(M,\widetilde{\Delta}_{d})|
\widehat{\mathcal{P}}^{2} | \psi_{0}(M,\widetilde{\Delta}_{d+\delta
  d})\rangle,
\end{eqnarray}
where $\delta\phi=\delta d/(1+d^{2})$ and $\delta
\Delta=-\frac{1}{2}\Delta (1+d^{2})^{-3/2}
\delta d$, and $\chi_{\Delta}$ is the FS with respect to the
anisotropy,  
\[
\chi_{\Delta}= \lim_{\delta\Delta\to 0} \frac{1 - |F(\Delta,\delta\Delta)|^2}{\delta\Delta^{2}}.
\]
 $\chi_{\Delta}$ scales with the system size $L$ in
a standard way~\cite{Yang}, i.e., linearly, so it can be neglected if we are
interested only in the leading $L$ dependence of $\chi_{J}$ which, as
we can already guess, is quadratic.

We will further assume that the total $z$-projection of the spin
$S^{z}_{\text{tot}}=\sum_j S_j^z=ML$ is a good quantum number, then
even at finite magnetization the ground state wave function
$|\psi_{0}(M,\widetilde{\Delta})\rangle$ can be made real, so that the
geometric connection term
\[
\langle \psi_{0}(M,\widetilde{\Delta}_{d})| \widehat{\mathcal{P}} |
\psi_{0}(M,\widetilde{\Delta}_{d+\delta d}) \rangle - (\text{h.c.})
\]
vanishes. Then the leading term in the current FS can be written as 
\begin{equation}
\label{fs-variation}
\chi_{J}^{o}(d,M)\simeq\frac{1}{(1+d^{2})^{2}}\Big( \langle
\widehat{\mathcal{P}}^{2}\rangle - \langle
\widehat{\mathcal{P}}\rangle^{2}\Big),
\end{equation}
where the averages here and in what follows are  taken in the ground
state  $|\psi_{0}(M,\widetilde{\Delta}_{d})\rangle$. This in turn
leads to the formula
\begin{equation}
\label{fs-via-correlator}
\chi^{o}_{J}(d,M) 
=\frac{ \displaystyle\sum_{j,j'=1}^{L} (j- j')^2\Big(
 \langle S_j^z \rangle  \langle S_{j'}^z\rangle   - \langle S_j^z
S_{j'}^z\rangle   \Big)}{2(1+d^2)^2},
\end{equation}
where we have again used the assumption that $S^{z}_{\text{tot}}$ is
conserved and hence 
\[
\sum_{j,j'} j^2(  \langle S_j^z S_{j'}^z\rangle -  \langle S_j^z
\rangle  \langle S_{j'}^z\rangle   )=0.
\]
Note that the evaluation of the CFS is simplified drastically for open
boundary conditions, since it is reduced to the task of calculating
the spin-spin correlation functions in the ground state.

Only the smooth part of the correlation function $\langle S_j^z
S_{j'}^z\rangle$ contributes to the leading size dependence of
$\chi_{J}^{o}$. This smooth part has the following universal behavior
\cite{HikiharaFurusaki04} (see also Ref.\ \cite{MikeskaPesch77} where exact correlation
functions for an open spin-$\frac{1}{2}$ $XY$ chain have been calculated):
\begin{eqnarray}
\label{corfun}
 \langle S_j^z S_{j'}^z\rangle &-&  \langle S_j^z \rangle  \langle
S_{j'}^z\rangle  \nonumber\\
&=&-\frac{K}{2\pi^2} \Big[    \frac{1}{f_2(j-j')}+ \frac{1}{f_2(j+j')}
  \Big] + \cdots, \nonumber\\
f_2(x)&=&\left[  \frac{2(L+1)}{\pi } \sin{\frac{\pi x}{2(L+1)}  }
  \right]^2 .
\end{eqnarray}
where $(\cdots)$ denotes oscillating terms, and $K=K(M,\widetilde{\Delta})$ is the Luttinger
liquid parameter that depends on the effective anisotropy
$\widetilde{\Delta}$ and the magnetization per site $M$. For
$S=\frac{1}{2}$ and $M=0$, it
is given by
\begin{equation} 
\label{K-Delta} 
K(M=0,\widetilde{\Delta})=\frac{\pi}{2\arccos(-\widetilde{\Delta})},
\quad \widetilde{\Delta}=\frac{\Delta}{(1+d^{2})^{1/2}}.
\end{equation}

In the limit $L\to\infty$ one can transform the sums in
(\ref{fs-via-correlator}) into integrals. Introducing the relative and center of
mass coordinates, $r=j-j'$ and $R=(j+j')/2$, we get,
\begin{eqnarray}
\sum_{j,j'}\frac{(j-j')^2}{f_2(j-j')}&\to& 2\int_0^{L-r} \mathrm{d} R \int_0^L \mathrm{d}r
\frac{r^2}{f_2(r)} \nonumber\\
&=&2\int_0^L\mathrm{d}r \frac{(L-r)r^2}{f_2(r)} \nonumber\\
&=& \frac{4L^2}{\pi} \int_0^{\pi/2} \mathrm{d}y \frac{(1-2y/\pi)y^2}{\sin^2{y}}\nonumber\\
&=&\frac{4L^2}{\pi^2}[ -\frac{1}{2} \pi^2 \ln{2}+\frac{21}{4} \zeta(3) ]
\end{eqnarray}
and
\begin{eqnarray}
\sum_{j,j'}\frac{(j-j')^2}{f_2(j+j')} &\to& 2\int_0^{L/2} \mathrm{d} R \int_{-2R}^{2R} \mathrm{d}r
\frac{r^2}{f_2(2R)} \nonumber\\
&=& \frac{4}{3} \int_0^{L/2} \mathrm{d} R  \frac{(2R)^3}{f_2(2R)} \nonumber\\
&=&\frac{8L^2}{3\pi^2}\int_0^{\pi/2}\mathrm{d} y \frac{y^3}{\sin^2{y}}\nonumber\\
&=& \frac{L^2}{\pi^2}(2\pi^2 \ln{2}-7\zeta(3)).
\end{eqnarray}

We obtain the final
result for the system size dependence of the current FS in the gapless spin-$\frac{1}{2}$
XXZ chain with open boundaries, in the following form: 
\begin{equation}
\label{open-CFS-spin-general}
\frac{\chi^{o}_{J}(d,M)}{L^2}= \frac{7\zeta(3)}{2\pi^4 (1+d^2)^2} 
K(M,\widetilde{\Delta})  +O(\frac{1}{L}).
\end{equation}
For $S=\frac{1}{2}$ chain at zero magnetization,  one can use the formula (\ref{K-Delta})
for the LL parameter to obtain a closed expression for the CFS; it is
easy to see that $\chi_{J}^{o}(d=0,M=0)$ has a singular behavior at $\Delta=1$.

The $L^2$ dependence of the current FS is a generic feature for
gapless models with o.b.c. and 
conserved $S^z$, where one can eliminate the current term (the DM
interaction) by means of a unitary ``twist'' operator  (\ref{twist}),
and where the smooth part
of the $\langle S^z_{j} S^z_{j'}\rangle$ correlator decays like $1/|j-j'|^2$.

\subsection{Relation to the tilt fidelity susceptibility}
\label{subsec:spintilt}

In a spin chain with open boundaries,  one can study another quantity,
which is, as we will show, related to the current FS, namely, the fidelity susceptibility
$\chi_{\mathcal{P}}$ with respect to
the ``polarization'' operator $\widehat{\mathcal{P}}=\sum_{j}jS^{z}_{j}$. For a spin-$\frac{1}{2}$
chain, this physically means a response to the gradient of the
external magnetic
field. For  the equivalent system of spinless fermions this could be a response to the
the ``lattice tilt'', or, if one assumes that the particles are
eletrically charged, then this is a response to the external electric field.
The tilt FS is given by
\begin{equation}
\label{tilt-fs}
\chi_{\mathcal{P}}= \sum_{n\not=0} \frac{ |\langle \psi_0|\widehat{\mathcal{P}} |\psi_n\rangle|^2} {(E_n-E_0)^2}.
\end{equation}
It is easy to see that the tilt FS is related to the ``dynamic polarizability'' $\alpha(\omega)$
\begin{equation}
\label{alpha}
\alpha(\omega)= \frac{\pi}{L}  \sum_{n\not=0}   |\langle
\psi_0|\widehat{\mathcal{P}}  
|\psi_n\rangle|^2 \delta(\omega-(E_n-E_0))
\end{equation}
by the following formula:
\begin{equation}
\label{tilt-fs-via-alpha}
\chi_{\mathcal{P}}=\frac{ L}{\pi}\int_{0}^{\infty} d\omega   \frac{ \alpha(\omega)}{\omega^2 }. 
\end{equation}
On the other hand,  one has 
\[
i[\widehat{H}_{\rm XXZ},
  \widehat{\mathcal{P}}]=\sum_l l(J_l-J_{l+1})=\widehat{J},
\]
and thus $\sigma_1({\omega})=\omega \alpha(\omega)$, which leads to
the following relation between the tilt FS and the conductivity:
\begin{equation}
\chi_{\mathcal{P}}= \frac{ L}{\pi}\int_{0}^{\infty} d\omega  \frac{ \sigma_1(\omega)}{\omega^3 } . 
\end{equation}
Thus, the leading term in the finite size scaling of
$\chi_{\mathcal{P}}$, similarly to the CFS $\chi_{J}$, will be
determined just by the low-frequency behavior of the conductivity.
Using the formulas for the conductivity (\ref{drude+regular}) and
(\ref{sigma-int2}), one arrives at the following result:
\begin{equation} 
\label{TFS-gen} 
\chi_{\mathcal{P}} =\frac{31KL^4 \zeta(5)}{8 u^2 \pi^6}.
\end{equation}

For free fermions ($\Delta=0$), the above result can be also reproduced directly in the same
way as it has been done in Eqs.\ (\ref{cond2})-(\ref{sigma-free}) for
the conductivity, using the ``density of states'' (\ref{dos}) and the
explicit expression for the matrix element (see the Appendix),
\begin{equation}
\label{matel-P}
 |\langle \psi_0|\widehat{\mathcal{P}}  |\psi_{m}\rangle|=\frac{L\big[ 1-(-1)^m\big]}{(m\pi)^2}.
\end{equation}
Indeed, using the perturbative expression
\begin{equation}
\chi_{\mathcal{P}}= \sum_{m>0} \rho(m) \frac{| \langle \psi_0|\widehat{\mathcal{P}}  |\psi_{m}\rangle  |^{2}} {(E_m-E_0)^2}
\end{equation}
with the linearized spectrum $E_{m}-E_{0} \simeq {u \pi m }/{L}$, one
obtains for free fermions
\begin{eqnarray}
\label{TFS-free}
\chi_{\mathcal{P}} (\Delta=0) &=& \sum^{\infty}_{k=0}\frac{4(2k+1)  L^2 }{ (E_{2k+1}-E_0)^2 (2k+1)^4\pi^4 }\nonumber\\
&=&\frac{31L^4 \zeta(5)}{8u^2 \pi^6}.
\end{eqnarray}
For interacting fermions $S^z$ gets the additional factor of $\sqrt{K}$ (see the Appendix),
hence bringing us back to the general result (\ref{TFS-gen}).

For the ratio of the current FS and the tilt FS in open chains one obtains the
universal result
\begin{equation}
\frac{L^2 \chi_J}{\chi_{\mathcal{P}}}=\frac{28u^2\pi^2 \zeta(3)}{31\zeta(5)}\simeq 10.3341 u^2.
\end{equation}

\section{Current FS in the fermionic Hubbard model}
\label{sec:FHM}

\begin{figure}[tb]
\includegraphics[width=6.5cm]{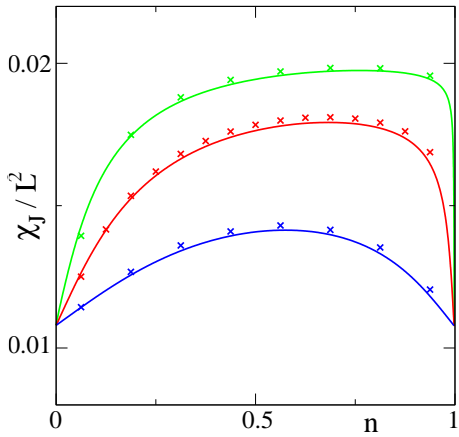}
\caption{ (Color online). Current fidelity susceptibility of the fermionic Hubbard
  model with o.b.c. for $U/t=1$, $2$ and $6$ (top to bottom). 
Symbols denote the DMRG data for Hubbard chains of $L=128$ sites. The lines correspond to the analytical
expression (\ref{Hubbardresult}). }
\label{fig:Hubbard}
\end{figure}

Consider the Hubbard model for spin-$\frac{1}{2}$ fermions (attractive or
repulsive), at arbitrary filling:
\begin{equation}
\widehat{H}_{\rm 0}=-\sum_{j,\sigma}
\left(c^{\dagger}_{j,\sigma}c^{\vphantom{\dagger}}_{j+1,\sigma}+\text{H.c.}\right)
+U\sum_{j}n^2_i,
\end{equation}
where $c_{j,\sigma}$ annihilates a fermion at site $j$ with the spin
$\sigma=\{\uparrow, \downarrow\}$,
 and $n_j=\sum_{\sigma}c^{\dagger}_{j,\sigma}c_{j,\sigma}$ is the
  fermion density at the site.
We assume open boundary conditions, and study the FS  $\chi_{J}^{o}$ with respect to
the total current, $\widehat{H}=\widehat{H}_{0}+\lambda
\widehat{J}_{\rm tot}$, with
\begin{equation}
\widehat{J}_{\rm tot}=-i \sum_{j,\sigma} \left(
c^{\dagger}_{j,\sigma}c^{\vphantom{\dagger}}_{j+1,\sigma}
-c^{\dagger}_{j+1,\sigma}c^{\vphantom{\dagger}}_{j,\sigma} \right).
\end{equation} 
The CFS can be calculated using the method of the unitary twist
operator, as described in  Sect.\ \ref{subsec:alt} for spin chains,
with the replacement of $S_{j}^{z}$ by
$n_j$.
One can closely follow all the steps of the calculation presented
above for a spin chain, and express the CFS through the density-density
correlation function of the Hubbard model. Assuming that its smooth
part has the form similar to  Eq.~(\ref{corfun}), with $K$ now being
the charge Luttinger parameter $K_c$ of the Hubbard model,
 we obtain the leading term in the finite-size scaling of the CFS as follows:
\begin{equation}
\label{Hubbardresult}
\frac{\chi^o_J}{L^2}= \frac{7\zeta(3)}{\pi^4} K_c(\nu,M)   + O(\frac{1}{L}),
\end{equation}
where $\nu$ is the lattice filling and $M$ is the magnetization.
Fig.\ \ref{fig:Hubbard} shows the theoretical curve corresponding to
Eq.~(\ref{Hubbardresult}) for the repulsive Hubbard model at $M=0$, versus
numerical results obtained by means of the density matrix renormalization group
(DMRG) technique for open chains of up to $L=128$ sites. The agreement between
the analytical expression and numerical results is quite good, especially taking
into account the fact that our analytical result (\ref{Hubbardresult}) concerns
only the $\sim L^2$ contribution.

Similar to the case of spin chains, one can study the tilt FS  $\chi_{\mathcal{P}}$ (i.e., the
response to the perturbation determined by $\widehat{W}=\sum_j j n_j
$) of the fermionic Hubbard
model with gapless charge excitations. Physically, such a perturbation
can be either the lattice tilt (for atoms in optical lattices), or
simply the external electric field (for charged particles).
Proceeding in a close analogy to
Sec.\ \ref{subsec:spintilt}, we obtain
\begin{equation}
\label{Hubbard-tilt}
\chi_{\mathcal{P}} =\frac{31K_cL^4 \zeta(5)}{4 u_c^2 \pi^6}.
\end{equation}

Finally, a few remarks are in order concerning the behavior of Hubbard
chains with p.b.c. One can again use the general connection between
the current FS and the conductivity, as we have done for spin chains,
but now it is the charge current and the charge conductivity, respectively.
In the repulsive Hubbard model at half-filling and  at any magnetization, the
charge excitations are gapped, so we expect the linear scaling of CFS independent of boundary conditions. 
Away from the half-filling, using the low frequency result for the conductivity
of doped Mott insulators $ \sigma_{\rm reg}(\omega) \sim \omega^3$
\cite{GiamarchiMillis}, one again obtains a linear
finite-size scaling, $\chi^p_J \sim L$, for any filling $\nu$ and
magnetization $M$. The same behavior (linear scaling of the CFS) we expect for
the attractive case away from half-filling as well, at any filling and
magnetization. 

At half filling, however, 
provided the perturbative result \cite{Giamarchi91}
$ \sigma_{\rm reg}(\omega) \sim U^2 \omega^{4K_c-5}$ holds  for the Hubbard
model, one obtains
\begin{eqnarray} 
\label{Hubbard-per-attr} 
&& \chi^p_J \sim L,\quad K_c>5/4,\nonumber\\
&& \chi^p_J \sim L^{6-4K_c},\quad K_c<5/4.
\end{eqnarray}
The point $K_c=1$ is special in Hubbard model, as it corresponds to $U=0$ and
hence $\chi^p_J=0$ there.

\section{Current FS in gapped systems}
\label{sec:gapped}

Up to now, we have dealt with systems that have gapless excitation spectrum, apart of the comment on Hubbard model at half-filling in the previous section. It
is easy to see that the system size dependence of the CFS in gapped systems is
generically linear, $\chi_{J}\propto L$, independent of
boundary conditions. The reason is that Drude part disappears in gapped phases and conductivity vanishes at energies below excitation gap $\omega_0$ (no excited states are available below gap), $\sigma_1(\omega)\sim \Theta(\omega-\omega_0)$, where $\Theta(x)$ is a step function. Alternatively for systems with o.b.c., the unitary transformation approach of
Sec.\ \ref{subsec:alt} (which is applicable in case of a pure chain geometry,
i.e., in absence of next-nearest-neighbor and longer range hoppings) can be
utilized for gapped spin chains as well, and leads to the formulas
(\ref{fs-variation}) and (\ref{fs-via-correlator}) connecting the CFS and the
reduced longitudinal spin-spin correlator. In a gapped system (for example, in
the N\'eel state of the spin-$\frac{1}{2}$ XXZ chain at $\Delta>1$, or in the
N'eel and rung-singlet phases of the spin-$\frac{1}{2}$ XXZ ladder, see below), this correlator
decays exponentially, so the sum in (\ref{fs-via-correlator}) will be
proportional to $L$.  A similar argument can be applied for fermionic or bosonic models.

Numerically, if the gap is extremely small, it may be
difficult to distinguish exponential decay from algebraic one; for the FS this
would mean distinguishing the linear scaling $\chi_{J}\propto CL$ with a large
prefactor $C$ from the quadratic scaling, $\chi_{J}\propto L^{2}$.

\begin{figure}[tb]
\includegraphics[width=6.5cm]{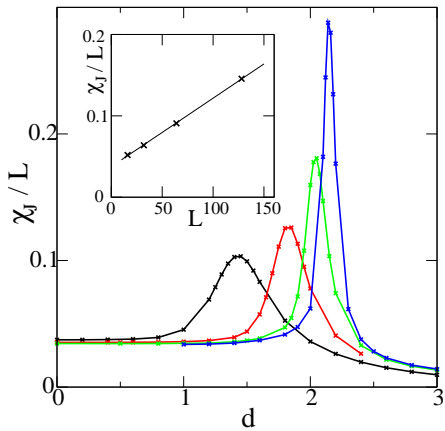}
\caption{ The CFS of a spin-$\frac{1}{2}$ AFM ladder defined by (\ref{H-lad}),
  with o.b.c., in the vicinity of the Ising phase transition from the N\'eel to
  the rung-singlet state, 
  for $J_R=3$ and $\Delta=1.5$ and system sizes $L=16$, $32$, $64$ and $128$ rungs. The inset shows that the peak of the CFS per
  site at the transition point scales linearly with the system size. }
\label{fig:Ladder}
\end{figure}

We illustrate the generic behavior of the CFS in gapped systems on the example
of the spin-$\frac{1}{2}$ antiferromagnetic spin ladder defined by the Hamiltonian
\begin{eqnarray}
\label{H-lad}
\widehat{H}_{\text{Lad}}&=& \sum_{l,\alpha}\big[ S^x_{l,\alpha}S^x_{l+1,\alpha }+S^y_{l,\alpha}S^y_{l+1,\alpha}
+\Delta S^z_{l,\alpha}S^z_{l+1,\alpha}  \big]\nonumber\\ 
&+&J_R\sum_{l} \vec{S}_{l,1}\cdot\vec{S}_{l,2} 
+d \sum_{l,\alpha}(\vec{S}_{l,\alpha}\times \vec{S}_{l+1,\alpha})^z ,
\end{eqnarray}
where $\alpha=1,2$ denotes the two legs of the ladder.  In
Fig.\ \ref{fig:Ladder}, we show the DMRG results for the CFS in the vicinity of
the Ising quantum phase transition between the N\'eel and rung-singlet
states. Ordinary quadratic scaling of the CFS peak at transition and linear
scaling of the wings is observed.

However, there are peculiar cases when the CFS may have a nontrivial finite size
scaling in a gapped system with open boundaries. Namely, in a topologically
ordered system, the presence of entangled edge spins localized at the boundaries
may render the sum in (\ref{fs-via-correlator}) $\sim L^2$, despite the
exponentially decaying correlation function. Let us take the spin-$1$ Haldane
chain as an example. The topologically ordered
\cite{denNijsRommelse89,GirvinArovas89} ground state of the open Haldane chain
is ``nearly'' fourfold degenerate \cite{KennedyTasaki92} due to the presence of
spin-$\frac{1}{2}$ edge spins localized at the boundaries: the lowest state is a
singlet, which is split from the Kennedy-Tasaki triplet by the exponentially
small ``boundary gap'' $\propto e^{-L/\xi}$, where $\xi\sim 6$ is the bulk
correlation length. In the singlet ground state, the reduced correlator between
the edge spins remains finite, so, according to (\ref{fs-via-correlator}),
$\chi_{J}\propto L^{2}$. It is clear that this behavior will be typical for any
state characterized by the presence of edge spins that are non-locally
entangled with each other.

The scaling of $\chi_{J}$ will be very sensitive to the numerical errors; for
example, if one accidentally takes a non-entangled member $|\uparrow
\uparrow\rangle$ of the Kennedy-Tasaki triplet as the ground state, the reduced
correlator between the edge spins will become zero, resulting in the generic
linear behavior $\chi_{J}\propto L$.

\section{Summary}
\label{sec:summary}

Combining numerical simulations with analytical arguments based on bosonization,
we have studied the finite size scaling of the current fidelity susceptibility
$\chi_{J}$ with respect to the charge or spin current in gapless one-dimensional
lattice models. We related it to the low-frequency behavior of the corresponding
conductivity, and identified the main reason for different scaling laws in
systems with open and periodic boundary conditions with the absence of the zero
mode of the current operator for the former case. For systems with p.b.c.
$\chi_{J}$ is directly connected to the low frequency behavior of the
\emph{regular} part of the conductivity, while in open systems $\chi_{J}$ is
determined by the \emph{singular} part of the conductivity that is essentially
the smeared Drude peak.

For the systems with o.b.c. we obtained the universal quadratic scaling
$\chi_{J}\propto L^2$, which obscurs the detection of quantum phase transitions
between two gapless regions from the finite-size scaling of the peak in
$\chi_{J}$.  Furthermore, for open chains we related $\chi_{J}$ with the
``tilt'' fidelity susceptibility that describes the response to the gradient of
the chemical potential.

In the future studies, it would be interesting to perform numerical calculations
of $\chi_{J}$ for large periodic spin-$\frac{1}{2} $$XXZ$ chains $L>100$, to confirm the
nontrivial low-frequency behavior of the regular conductivity predicted by
Giamarchi \cite{Giamarchi91} for $\Delta \lesssim 1$. It would also be
interesting to study 1d models with iTEBD method and determine the scaling of
$\chi_{J}$ with matrix dimension. Similar studies in higher dimensions can also be interesting.

\acknowledgments

We thank Eric Jeckelmann for helpful discussions. This work has been supported
by QUEST (Center for Quantum Engineering and Space-Time Research) and DFG
Research Training Group (Graduiertenkolleg) 1729.

\appendix
\section*{Appendix}

In this Appendix, we derive analytical expressions for matrix elements of the
total momentum $\int_0^L \tilde{\Pi}\, dx$ and the 'center-of-mass' operator
$\int_0^L x \partial_x \tilde{\phi}(x)\, dx$ between the vacuum and excited
states of the Gaussian bosonic model, for the case of zero boundary conditions.

We start with the total momentum operator. 
It is convenient to expand the bosonic fields in the Fourier modes of the open string,
\begin{eqnarray}
\tilde \phi(x)&=&\sqrt{\frac{2}{L}}\sum^{\infty}_{n=1}  
\sin\frac{\pi n x}{L}  \tilde \phi_n\nonumber\\
\tilde \Pi(x)&=&\sqrt{\frac{2}{L}}\sum^{\infty}_{n=1}  
\sin\frac{\pi n x}{L}  \tilde \pi_n,
\end{eqnarray}
which guarantees the boundary conditions $\tilde{\phi},\tilde {\Pi}=0$ at the
chain ends. The inverse relations,
\begin{eqnarray}
\tilde \phi_n & =&   \sqrt{\frac{2}{L}}
\int_0^L  \sin\frac{\pi n x}{L} \tilde \phi(x)\, dx \nonumber\\
 \tilde \pi_n & =&    \sqrt{\frac{2}{L}}
\int_0^L  \sin\frac{\pi n x}{L} \tilde \Pi(x) \,dx
\end{eqnarray}
imply that zero modes for open chain do not exist,
\begin{equation}
\label{nozeromodes}
\tilde \phi_0\equiv  0, \quad \tilde \pi_0\equiv 0.
\end{equation}
Commutation relations of the Fourier modes are canonical,
\begin{equation}
[\phi_n,\pi_m]=i\delta_{n,m}.
\end{equation}

The total momentum operator in terms of the Fourier components can be rewritten as
\begin{eqnarray}
\int_0^L \tilde \Pi\, dx&=& \int_0^L
\sqrt{\frac{2}{L}}\sum^{\infty}_{n=1}  \sin\frac{\pi n x}{L}  \tilde
\pi_n   \, dx\nonumber\\
&=&\sqrt{2L}\sum_{n=1}^{\infty}\frac{1-(-1)^n}{\pi n}  \tilde \pi_n ,
\end{eqnarray}
and the Luttinger liquid Hamiltonian reads
\begin{eqnarray}
\widehat{H}_{LL}&=&\frac{u}{2}\int_0^L \big[ (\partial_x \tilde \phi)^2+\tilde
  \Pi^2 \big] d x \nonumber\\
&=& \frac{u}{2}  \sum^{\infty}_{n=1}\left[\tilde{\pi}_n^2 + 
\Big(\frac{\pi n}{L}\Big)^2\tilde{\phi}_n^2 \right]\nonumber\\
 &=&\frac{\pi u}{L}\sum_{n>0} n \big[\tilde{a}^{\dagger}_n \tilde{a}_n
  +\frac{1}{2} \big]  
= \sum_{n>0}\omega_n \tilde a^{\dagger}_n \tilde a_n +E_0.\nonumber
\end{eqnarray}
Here 
\begin{eqnarray}
 \tilde a^{\dagger}_n&=&\sqrt{ \frac{L}{2 \pi n}} \tilde \pi_n+i \sqrt{\frac{\pi n}{ 2L}} \tilde \phi_n \nonumber\\
 \tilde a_n&=&\sqrt{ \frac{L}{2 \pi n}} \tilde \pi_n-i \sqrt{\frac{\pi n}{ 2L}} \tilde \phi_n
\end{eqnarray}
are the standard bosonic creation and annihilation operators, $\tilde
a_n|0\rangle=0$, satisfying the commutation relations
$[a_n,a^{\dagger}_m]=\delta_{n,m}$, and defining the eigenstates $\widehat{H}_{LL}\tilde
a^{\dagger}_n|0\rangle=(\omega_n+E_0)\tilde a^{\dagger}_n|0\rangle $ .

The total momentum in this basis obtains the following form:
\begin{eqnarray}
\int_0^L \tilde \Pi\, dx=L\sum_n\frac{1-(-1)^n}{\sqrt{\pi n}}
\big[\tilde a_n+\tilde a^{\dagger}_n \big] .
\end{eqnarray}
Hence,
\begin{equation}
\langle n|   \int_0^L \tilde \Pi\mathrm{d} x |0\rangle =L\frac{1-(-1)^n}{\sqrt{ \pi  n}  }.
\end{equation}
The matrix elements of the fermionic total current are obtained from the relation,
\[
\widehat{J} \to \frac{ \sqrt{K}u}{\sqrt{\pi}} \int_0^L  { \tilde \Pi}\, d x. 
\]
so that,
\begin{eqnarray}
\label{dens}
\rho(n) |\langle \psi_n|\widehat{J}  |\psi_{0}\rangle|^2 
&=&\frac{ K  u^2}{{\pi}} |\langle n|   \int_0^L \tilde \Pi\mathrm{d} x |0\rangle|^2  \nonumber\\
&=&   \frac{K L^{2} u^2 \big[1-(-1)^n\big]^{2}}{\pi^2 n}.
\end{eqnarray}
Note that in the bosonized formulation each eigenstate with the energy $E_n=u\pi
n/L+E_0$, obtained from the vacuum by acting with the total momentum operator,
involves a single state $a^{\dagger}_n|0\rangle$; in other words, the bosonic
density of states is $\rho_{\text{bosonic}}(n)=1$, as opposed to the fermionic
picture where $\rho(n)=n$.

In a similar way, we can calculate the matrix elements of the polarization
(center of mass)
operator 
\[
\widehat{\mathcal{P}}=\sum_{j}jS^{z}_{j} \mapsto 
-\sqrt{\frac{K}{\pi}}\int_0^L x \partial_x \tilde{\phi}(x)\, dx.
\]
One obtains
\begin{eqnarray}
-\int_0^L  x \partial_x  \tilde \phi (x) \mathrm{d} x&=& 
- \sum_n\frac{\sqrt{2L}}{\pi n} \int_0^{\pi n}y \cos{y} \mathrm{d} y \tilde \phi_n\nonumber\\
&=& \sqrt{2L}\sum _n \frac{1-(-1)^n}{\pi n} \tilde \phi_n .
\end{eqnarray}
On the other hand,
\begin{equation}
\tilde \phi_n =\sqrt{\frac{2L}{\pi n}}\frac{a^{\dagger}_n-a_n}{2i}
\end{equation}
hence, 
\begin{eqnarray}
-\int_0^L  x \partial_x  \tilde \phi (x) \mathrm{d} x= L\sum_n \frac{1-(-1)^n}{(\pi n)^{3/2}} \frac{a^{\dagger}_n-a_n}{i}.
\end{eqnarray}
In a full similarity to Eq.~(\ref{dens}), we arrive at the generalization of
Eq.~(\ref{matel-P}) to the interacting case:
\begin{equation}
\label{polar}
\rho(n) |\langle \psi_n|\widehat{\mathcal{P}}  |\psi_{0}\rangle|^2 
 =   \frac{K L^{2}\big[(1-(-1)^n)\big]^{2}}{\pi^4 n^{3}}.
\end{equation}


\end{document}